\documentclass[11pt]{article}
\usepackage{color,latexsym,amsmath,amssymb,path,enumerate,url}

\newcommand{\ignore}[1]{}

\newtheorem{theorem}{Theorem}[section]
\newtheorem{lemma}[theorem]{Lemma}
\newtheorem{corollary}[theorem]{Corollary}

\newcommand{\Proof}[1]
        {
        \noindent
        \emph{Proof #1.}~
        }
\newsavebox{\smallProofsym}                     
\savebox{\smallProofsym}                        %
        {
        \begin{picture}(7,7)                    %
        \put(0,0){\framebox(6,6){}}             %
        \put(0,2){\framebox(4,4){}}             %
        \end{picture}                           %
        }                                       %
\newcommand{\smalleop}[1]
        {
        \mbox{} \hfill #1~~\usebox{\smallProofsym}\!\!\!\!\!\!\
        }
\newenvironment{theProof}[1]
        {
        \Proof{#1}}{\smalleop{}
        \medskip

        }
\usepackage{graphicx,psfrag}
\usepackage[rflt]{floatflt}

\newcommand{\placefig}[2]
        {\includegraphics[width=#2]{#1.eps}}
\usepackage{dsfont}

\newcommand{\tri}[1]{{\mathsf{tr}}(#1)}
\newcommand{\btri}[1]{{\bf tr}(#1)}

\newcommand{\tr}[1]{{\cal T}(#1)}

\newcommand{\pg}[1]{{\cal P}(#1)}
\newcommand{\pgc}[2]{{\cal P}^=_{#2}(#1)}
\newcommand{\pgcc}[2]{{\cal P}^\le_{#2}(#1)}
\newcommand{\pgccc}[2]{{\cal P}^\ge_{#2}(#1)}

\newcommand{\st}[1]{{\cal ST}(#1)}
\newcommand{\f}[1]{{\cal F}(#1)}

\newcommand{\pgr}[1]{{\mathsf{pg}}(#1)}
\newcommand{\pgrc}[2]{{\mathsf{pg}^=_{#2}}(#1)}
\newcommand{\pgrcc}[2]{{\mathsf{pg}^\le_{#2}}(#1)}
\newcommand{\pgrccc}[2]{{\mathsf{pg}^\ge_{#2}}(#1)}
\newcommand{\str}[1]{{\mathsf{st}}(#1)}
\newcommand{\fr}[1]{{\mathsf{f}}(#1)}

\newcommand{\kfor}[2]{{\mathsf{f}_{#2}}(#1)}

\newcommand{\supp}[1]{\mathsf{supp}(#1)}
\newcommand{\flip}[0]{\mathsf{flip}}

\begin{document}
\pagenumbering{arabic}
\date{}

\title{Counting Plane Graphs: Flippability and its Applications\thanks{%
Work on this paper by Micha Sharir and Adam Sheffer was partially supported by Grant 338/09 from
the Israel Science Fund. Work by Micha Sharir was also
supported by NSF Grant CCF-08-30272,
by Grant 2006/194 from the U.S.-Israel Binational Science Foundation,
and by the Hermann
Minkowski--MINERVA Center for Geometry at Tel Aviv University.
Work by Csaba D. T\'oth was supported in part by NSERC grant RGPIN 35586. Research by this
author was conducted at ETH Z\"urich.
Emo Welzl acknowledges support from the EuroCores/EuroGiga/ComPoSe SNF grant 20GG21\_134318/1.
Part of the work on this paper was done at the Centre Interfacultaire Bernoulli (CIB), EPFL, Lausanne,
during the Special Semester on Discrete and Computational Geometry, Fall 2010, and was supported by the Swiss National Science Foundation.}}

\author{
Michael Hoffmann\thanks{Institute of Theoretical Computer Science,
ETH Z\"urich, CH-8092 Z\"urich, Switzerland.
Email: \texttt{\{hoffmann,welzl\}@inf.ethz.ch}}
\and
Andr\'e Schulz\thanks{%
Institut f\"ur Mathematische Logik und Grundlagenforschung, Universit\"at M\"unster,
Germany.
Email: \texttt{andre.schulz@uni-muenster.de}}
\and
Micha Sharir\thanks{%
School of Computer Science, Tel Aviv University,
Tel Aviv 69978, Israel and Courant Institute of Mathematical
Sciences, New York University, New York, NY 10012, USA.
Email: \texttt{michas@tau.ac.il}}
\and
Adam Sheffer\thanks{%
School of Computer Science, Tel Aviv University,
Tel Aviv 69978, Israel.
Email: \texttt{sheffera@tau.ac.il}}
\and
Csaba D. T\'oth\footnote{Department of Mathematics and Statistics,
University of Calgary, Calgary, AB, Canada\@. Email: \texttt{cdtoth@ucalgary.ca}.}
\and
Emo Welzl\footnotemark[2]
}

\maketitle

\begin{abstract}
We generalize the notions of flippable and simultaneously flippable edges in a triangulation of a set $S$ of points in the plane
to so-called \emph{pseudo-simultaneously flippable edges}. Such edges are related to the notion of convex decompositions spanned by $S$.

 We prove a worst-case tight lower bound for the number of pseudo-simultaneously flippable edges in a triangulation in terms of the number of vertices. We use this bound for deriving new upper bounds for the maximal number of crossing-free straight-edge graphs that can be embedded on any fixed set of $N$ points in the plane. We obtain new upper bounds for the number of spanning trees and forests as well.
 Specifically, let $\tri{N}$ denote the maximum number of triangulations on a set of $N$ points in the plane. Then we show (using the known bound $\tri{N} < 30^N$) that any $N$-element point set admits at most $6.9283^N \cdot \tri{N} < 207.85^N$ crossing-free straight-edge graphs, $O(4.7022^N) \cdot \tri{N} = O(141.07^N)$ spanning trees, and $O(5.3514^N) \cdot \tri{N} = O(160.55^N)$ forests.
We also obtain upper bounds for the number of crossing-free straight-edge graphs that have $cN$, fewer than $cN$, or more than $cN$ edges, for any constant parameter $c$, in terms of $c$ and $N$.
\end{abstract}

\section{Introduction}

A {\em crossing-free straight-edge graph} $G$ is an embedding of a planar graph in the plane such that the vertices are mapped to a set $S$ of points in the plane and the edges are pairwise non-crossing line segments between pairs of points in $S$. (Segments are allowed to share endpoints.) By F{\'a}ry's classical result \cite{FA48}, such an embedding is always possible. In this paper, we fix a labeled set $S$ of points in the plane, and we only consider planar graphs that admit a straight-edge embedding with vertex set $S$.  By \emph{labeled} we mean that each vertex of the graph has to be mapped to a unique designated point of $S$. Analysis of the number of plane embeddings of planar graphs in which the set of vertices is not restricted to a specific embedding, or when the vertices are not labeled, can be found, for example, in \cite{GN09,Mul65,Tut63}.

A \emph{triangulation} of a set $S$ of $N$ points in the plane is a
maximal crossing-free straight-edge graph on $S$ (that is, no additional straight edges can be inserted without crossing some of the existing edges).
Triangulations are an important geometric construct which is used in many algorithmic applications, and are also an interesting object of study in discrete and
combinatorial geometry (recent comprehensive surveys can be found in \cite{DRS10,HD09}).

Improving the bound on the maximum number of triangulations that any set of $N$ points in the plane can have has been
a major research theme during the past 30 years.
The initial upper bound $10^{13N}$ of \cite{ACNS82} has been steadily improved in several paper (e.g., see \cite{DeSo97,SaSe03,SW06}), culminating with the current record of $30^N$
due to Sharir and Sheffer \cite{SS10}. Other papers have studied lower bounds on the maximal number of triangulations (e.g., \cite{AHHHKV07,DSST11}), and upper or lower bounds on the number of other kinds of planar graphs (e.g., \cite{BKKSS07,BS10,Rib05,Rot05}).

Every triangulation of $S$ contains the edges of the convex hull of $S$, and the remaining edges of the triangulation decompose the interior of the convex hull into triangular faces.
Assume that $S$ contains $N$ points, $h$ of which are on the convex hull boundary and the remaining $n=N-h$ points are interior to the hull (we use this notation throughout).
By Euler's formula, every triangulation of $S$ has $3n + 2h - 3$ edges ($h$ {\em hull} edges, common to all triangulations, and $3n + h - 3$ {\em interior} edges, each adjacent to two triangles), and $2n + h - 2$ bounded triangular faces.

\begin{figure}[h]
\centerline{\placefig{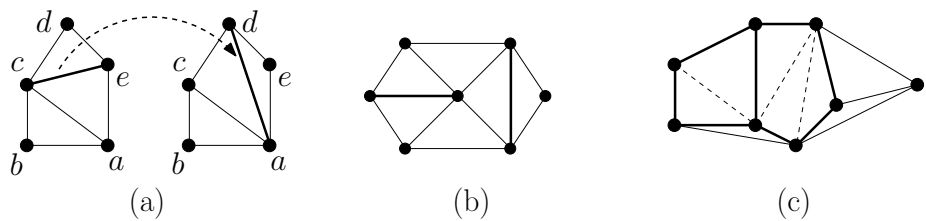}{0.8\textwidth}}
\vspace{-1mm}

\caption{\small \sf (a) The edge $ce$ can be flipped to the edge $ad$.
(b) The two bold edges are simultaneously flippable.
(c) Interior-disjoint convex quadrilateral and convex pentagon in a triangulation.}
\label{fi:intro}
\vspace{-2mm}
\end{figure}

\paragraph{Edge Flips.}
Edge flips are simple operations that replace one or several edges of a triangulation with new edges and produce a new triangulation.
As we will see in Section \ref{se:apps}, edge flips are instrumental for counting various classes of subgraphs in triangulations.
In the next few paragraphs, we review previous results on edge flips, and propose a new type of edge flip.
We say that an interior edge in a triangulation of $S$ is {\em flippable}, if its two adjacent triangles form a convex quadrilateral.
A flippable edge can be {\em flipped}, that is, removed from the
graph of the triangulation and replaced by the other diagonal of the corresponding quadrilateral, thereby obtaining a new triangulation of $S$.
An edge flip operation is depicted in Figure 1(a), where the edge $ce$ is flipped to the edge $ad$.
Already in 1936, Wagner \cite{Wagner36} has shown that any \emph{unlabeled abstract} triangulation $T$ (in this case, two triangulations are considered identical if we can relabel and change the planar embedding of the vertices of the first triangulation, to obtain the second triangulation) can be transformed into any other triangulation $T'$ (with the same number of vertices)
through a series of edge-flips (here one uses a more abstract notion of an edge flip).
When we deal with a pair of triangulations over a specific common (labeled) set $S$ of points in the plane, there always exists such a sequence of $O(|S|^2)$ flips,
and this bound is tight in the worst case (e.g., see \cite{BH09,HNU99}).
Moreover, there are algorithms that perform such sequences of flips
to obtain some ``optimal" triangulation (typically, the Delaunay triangulation; see \cite{F92} for example), which, as a by-product, provide an edge-flip
sequence between any specified pair of triangulations of $S$.

How many flippable edges can a single triangulation have? Given a triangulation $T$, we denote by $\flip(T)$
the number of flippable edges in $T$. Hurtado, Noy, and Urrutia \cite{HNU99} proved the following lower bound.
\begin{lemma} \label{le:flipOriginal} \emph{\bf \cite{HNU99}} For any triangulation $T$ over a set of $N$ points in the plane,
\[ \flip(T) \ge N/2 - 2. \]
Moreover, there are triangulations (of specific point sets of arbitrarily large size) for which this bound is tight.
\end{lemma}

\begin{figure}[h]

\centerline{\placefig{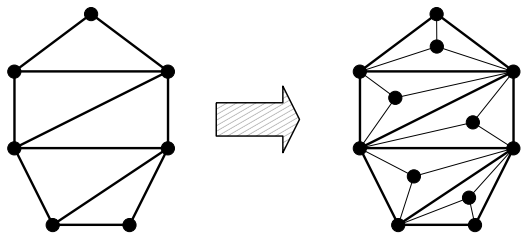}{0.45\textwidth}}
\vspace{-1mm}

\caption{\small \sf Constructing a triangulation with $N/2-2$ flippable edges.}
\label{fi:lowFlip}
\vspace{-2mm}
\end{figure}

To obtain a triangulation with exactly $N/2-2$ flippable edges (for an even $N$), start with a convex polygon with $N/2+1$ vertices, triangulate it in some arbitrary manner, insert
a new point into each of the $N/2-1$ resulting bounded triangles, and connect each new point $p$ to the three hull vertices that form the triangle
containing $p$. Such a construction is depicted in Figure \ref{fi:lowFlip}. The resulting graph is a triangulation with $N$ vertices and exactly $N/2-2$ flippable edges,
namely the chords of the initial triangulation.

Next, we say that two flippable edges $e$ and $e'$ of a triangulation $T$ are \emph{simultaneously flippable} if no triangle of $T$ is incident to both edges;
equivalently, the quadrilaterals corresponding to $e$ and $e'$ are interior-disjoint.
See Figure \ref{fi:intro}(b) for an illustration.
Notice that flipping an edge $e$ cannot affect the flippability of any edge simultaneously flippable with $e$. Given a triangulation $T$, let $\flip_s(T)$
denote the size of the largest subset of edges of $T$, such that every pair of edges in the subset is simultaneously flippable. The following lemma, improving upon an earlier weaker bound in \cite{GHNPU03}, is taken from
Souvaine, T\'oth, and Winslow \cite{STW11}.
\begin{lemma}
\label{le:sFlipOriginal} \emph{\bf \cite{STW11}}
For any triangulation $T$ over a set of $N$ points in the plane,
$\flip_s(T) \ge (N-4)/5$.
\end{lemma}
Galtier et al.~\cite{GHNPU03} show that this bound is tight in the worst case,
by presenting a specific triangulation in which at most $(N-4)/5$ edges are simultaneously flippable.

\paragraph{Pseudo-simultaneously flippable edge sets.} A set of simultaneously flippable edges in a triangulation $T$ can be considered as the set of   diagonals of a collection of interior-disjoint convex quadrilaterals.
We consider a more liberal definition of simultaneously flippable edges, by taking, within a fixed triangulation $T$, the diagonals of a set of interior-disjoint convex polygons, each with at least
four edges (so that the boundary edges of these polygons belong to $T$).
Consider such a collection of convex polygons $Q_1,\ldots ,Q_m$, where $Q_i$ has $k_i \ge 4$ edges, for $i=1,\ldots,m$. We can then retriangulate each $Q_i$ independently, to obtain many different triangulations. Specifically, each $Q_i$ can be triangulated in $C_{k_i-2}$ ways, where $C_j$ is the
$j$-th \emph{Catalan number} (see, e.g., \cite[Section 5.3]{St99}). Hence, we can get $M = \prod _{i=1}^{m}C_{k_i - 2}$ different triangulations in this way.
In particular, if a graph $G \subseteq T$ (namely, all the edges of $G$ are edges of $T$) does not contain any diagonal of any $Q_i$ (it may contain boundary edges though) then $G$ is a subgraph of (at least) $M$ distinct triangulations. An example is depicted in Figure \ref{fi:intro}(c), where by ``flipping" (or rather, redrawing) the diagonals of the highlighted quadrilateral and pentagon, we can get $C_2 \cdot C_3 = 2 \cdot 5 = 10$ different triangulations (including the one shown), and any subgraph of the triangulation that does not contain any of these diagonals is a subgraph of these ten triangulations.
We say that a set of interior edges in a triangulation is {\em pseudo-simultaneously flippable} ({\em ps-flippable} for short) if after the deletion of these edges every bounded face of the remaining graph is convex, and there are no vertices of degree 0. Notice that all three notions of flippability are defined within a fixed triangulation $T$ of $S$ (although each of them gives a recipe for producing many other triangulations).

\begin{table}[ht]
\centering
\caption{\small \sf Bounds for minimum numbers of the various types of flippable edges in a triangulation of $N$ points.
All of these bounds are tight in the worst case.} \label{ta:Flippable} \vspace{2mm}

\begin{tabular}{l | c }
\hline\hline
Edge Type & Lower bound  \\

\hline

Flippable &  $N/2 - 2$ \cite{HNU99}\\ [0.5ex]
Simultaneously flippable & $N/5 - 4/5$  \cite{GHNPU03,STW11} \\ [0.5ex]
Ps-flippable  & $\max\{N/2-2,h-3\}$ \\ [1ex]
\hline
\end{tabular}
\end{table}

\paragraph{Our results.} In Section \ref{sec:psFlip}, we derive a lower bound on the size of the largest set of ps-flippable edges in a triangulation,
and show that this bound is tight in the worst case. Specifically, we have the following result.
\begin{lemma}[ps-flippability lemma] \label{le:psFlip}
Let $S$ be a set of $N$ points in the plane, and let $T$ be a triangulation of $S$.
Then $T$ contains a set of at least $\max\{N/2-2,h-3\}$ ps-flippable edges. This bound is tight in the worst case.
\end{lemma}

Table \ref{ta:Flippable} summarizes the bounds for the minimum numbers of the various types of flippable edges in a triangulation.

\begin{figure}[h]

\centerline{\placefig{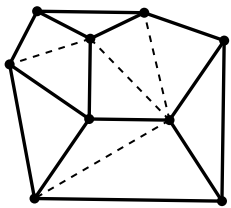}{0.2\textwidth}}
\vspace{-1mm}

\caption{\small \sf A convex decomposition of $S$. When completing it into a triangulation, the added (dashed)
diagonals form a set of ps-flippable edges. This is one of the $C_2\cdot C_2 \cdot C_3 = 20$ possible completions.}
\label{fi:decomposition}
\vspace{-2mm}
\end{figure}

We also relate ps-flippable edges to \emph{convex decompositions} of $S$. These are crossing-free straight-edge graphs on $S$
such that (i) they include all the hull edges, (ii) each of their bounded faces is a convex polygon, and (iii) no point of $S$ is isolated.
See Figure \ref{fi:decomposition} for an illustration.

\begin{table}[h]
\centering
\begin{tabular}{l | c | c | c | c}
\hline\hline
Graph type & Lower bound & Previous & New upper & In the form \\
                    &                    & upper bound  & bound     & $a^N \cdot \tri{N}$ \\

\hline
& & & & \\ [-2ex]

Plane Graphs & $\Omega(41.18^N)$ \cite{AHHHKV07} & $O(239.40^N)$ \cite{RSW08,SS10} & $\mathbf{207.85^N}$ & $\mathbf{6.9283^N\cdot\btri{N}}$\\ [0.5ex]
Spanning Trees & $\Omega(11.97^N)$ \cite{DSST11} & $O(158.56^N)$  \cite{BS10,SS10} & $\mathbf{O(141.07^N)}$ & $\mathbf{O(4.7022^N)\cdot\btri{N}}$\\ [0.5ex]
Forests & $\Omega(12.23^N)$ \cite{DSST11} & $O(194.66^N$) \cite{BS10,SS10} & $\mathbf{O(160.55^N)}$ & $\mathbf{O(5.3514^N) \cdot\btri{N}}$ \\ [1ex]
\hline
\end{tabular}
\caption{\small \sf Upper and lower bounds for the number of several types of crossing-free straight-edge graphs on a set of $N$ points in the plane.
By plane graphs, we mean all crossing-free straight-edge graphs embedded on a specific labeled point set.
Our new bounds are in the right two columns.} \label{ta:UpperBounds}
\end{table}
\paragraph{Counting plane graphs: New upper bounds.}
In Section \ref{se:apps}, we use Lemma \ref{le:psFlip} to derive several upper bounds on the numbers of planar graphs
of various kinds embedded as crossing-free straight-edge graphs on a fixed labeled set $S$.
For a set $S$ of points in the plane, we denote by $\tr{S}$ the set of all
triangulations of $S$, and put $\tri{S}:= \left|\tr{S}\right|$. Similarly, we denote by $\pg{S}$ the set of all
crossing-free straight-edge graphs on $S$, and put $\pgr{S}:= \left|\pg{S}\right|$. We also let $\tri{N}$ and $\pgr{N}$ denote,
respectively, the maximum values of $\tri{S}$ and of $\pgr{S}$, over all sets $S$ of $N$ points in the plane.

Since a triangulation of $S$ has fewer than $3|S|$ edges, the trivial upper bound
$\pgr{S} < 8^{|S|} \cdot \tri{S}$
holds for any point set $S$. Recently, Razen, Snoeyink, and Welzl \cite{RSW08} slightly improved the upper bound on the ratio $\pgr{S}/\tri{S}$ from $8^{|S|}$ down to $O\left(7.9792^{|S|}\right)$.
We give a more significant
improvement on the ratio with an upper bound of $6.9283^{|S|}$. Combining this bound with the recent bound $\tri{S} < 30^{|S|}$ \cite{SS10}, we get $\pgr{N}<207.85^N$.
We provide similar improved ratios and absolute bounds for the numbers of crossing-free straight-edge spanning trees and forests (i.e., cycle-free graphs). Table \ref{ta:UpperBounds} summarizes these results\footnote{Up-to-date bounds for these and for other families of graphs can be found in \url{http://www.cs.tau.ac.il/~sheffera/counting/PlaneGraphs.html} (version of November 2010).}.

\begin{figure}[h]

\centerline{\placefig{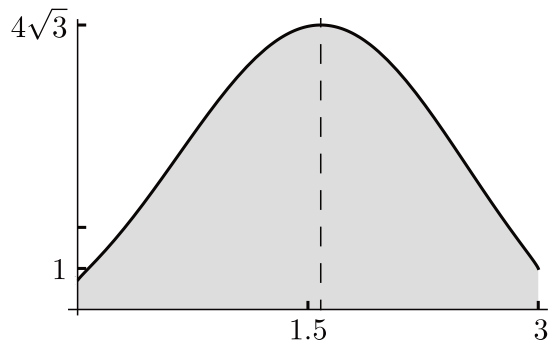}{0.47\textwidth}}
\vspace{-1mm}

\caption{\small \sf The base $B(c)$ of the exponential factor in the bound on the number of crossing-free straight-edge graphs with $c|S|$ edges, as a function of $c$. The maximum is attained at $c=19/12$ (see below).
}
\label{fi:plot1}
\vspace{-2mm}
\end{figure}

We also derive similar ratios for the number of crossing-free straight-edge graphs
with exactly $c|S|$ edges, with at least $c|S|$ edges, and with at most $c|S|$ edges, for $0 < c < 3$.
For the case of  crossing-free straight-edge graphs
with exactly $c|S|$  we obtain the bound\footnote{In the notations $O^*()$, $\Theta^*()$, and $\Omega^*()$, we neglect polynomial factors.} \[ O^*\left( \left( \frac{5^{5/2}}{8(c+t-1/2)^{c+t-1/2} (3-c-t)^{3-c-t}(2t)^t(1/2-t)^{1/2-t}} \right)^N \cdot \tri{S} \right), \]
where $t = \frac{1}{2} \left(\sqrt{(7/2)^2 + 3c + c^2 } - 5/2 - c \right)$. Figure \ref{fi:plot1} contains a plot of the base $B(c)$ of the exponential factor multiplying $\tri{N}$ in this bound, as a function of $c$.

\paragraph{Notation.}
Here are some additional notations that we use. \begin{list}{}{\leftmargin= 0em}
\item For a triangulation $T$ and an integer $i\geq 3$, let $v_i(T)$ denote the number of interior vertices of degree
$i$ in $T$. \vspace{-1.5mm}
\item Given two crossing-free straight-edge graphs $G$ and $H$ over the same point set $S$, we write $G \subseteq H$ to indicate that every edge in $G$ is also an edge in $H$.  \vspace{-1.5mm}
\item Similarly to the case of edges, the \emph{hull vertices} (resp., \emph{interior vertices}) of a set $S$ of points in the plane are those that are part of the boundary of the convex hull of $S$ (resp., not part of the convex hull boundary).
\item  We only consider point sets $S$ in general position, that is, no three points in $S$ are collinear. For upper bounds on the number of graphs, this involves no loss of generality, because the number of graphs can only increase if collinear points are slightly perturbed into general position.
\end{list}

\begin{figure}[h]

\centerline{\placefig{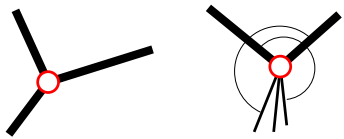}{0.3\textwidth}}
\vspace{-1mm}

\caption{\small \sf Separable edges.}
\label{fi:separable}
\vspace{-2mm}
\end{figure}

\noindent
\textbf{Separable edges.}
Let $p$ be an interior vertex in a convex decomposition $G$ of $S$. Following the notation in \cite{SSW10}, we call an edge $e$ incident
to $p$ in $G$  {\em separable at $p$} if it can be
separated from the other edges incident to $p$ by a line
through $p$ (see Figure \ref{fi:separable}, where the separating lines are not drawn). Equivalently, edge $e$ is separable at $p$ if the two angles between $e$ and its clockwise and
counterclockwise neighboring edges (around $p$) sum up to
more than $\pi$.
Following \cite{HNU99}, we observe the following easy properties, both of which materialize in Figure \ref{fi:separable}.
\begin{enumerate}[(i)]
\item If $p$ is an interior vertex of degree $3$ in $G$, its three incident edges are
separable at $p$, for otherwise $p$ would have been a reflex vertex of some face.
\vspace{-1.5mm}
\item An interior vertex $p$ of degree $4$ or higher can have at
most two incident edges which are separable at $p$ (and if it has two
such edges they must be consecutive in the circular order around $p$).
\end{enumerate}

\section{The size of ps-flippable edge sets} \label{sec:psFlip}

In this section, we establish the ps-flippability lemma (Lemma \ref{le:psFlip} from the introduction).
We restate the lemma for the convenience of the reader.
\vspace{-3mm}

\paragraph{Lemma \ref{le:psFlip}}\hspace{-2mm}{\bf (ps-flippability lemma)} \emph{Let $S$ be a set of $N$ points in the plane, and let $T$ be a triangulation of $S$.
Then $T$ contains a set of at least $\max\{N/2-2,h-3\}$ ps-flippable edges. This bound is tight in the worst case.}
\vspace{2mm}

\begin{theProof}{\!\!}
Starting with the proof of the lower bound, we apply the following iterative process to $T$.
As long as there exists an interior edge whose removal does not
create a non-convex face, we pick such an edge and remove it. When we
stop, we have a crossing-free straight-edge graph $G$, all of whose bounded faces are
convex; that is, we have a locally minimal convex decomposition of $S$.
Note that all $h$ original hull edges are still in $G$,
and that every interior vertex of $G$ has degree at least $3$ (recall the general position assumption).

\begin{figure}[h]
\centerline{\placefig{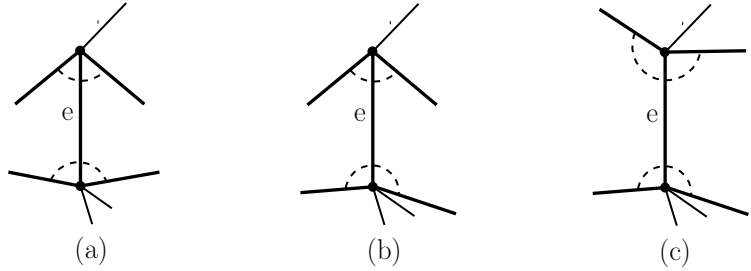}{0.65\textwidth}}
\vspace{-1mm}

\caption{\small \sf (a) An edge not separable at both of its endpoints can be removed from the graph.
   (b,c) An edge separable in at least one of its endpoints cannot be removed from the convex decomposition.}
\label{fi:nonessential}
\vspace{-2mm}
\end{figure}

Note that every edge of $G$ is separable at one or both of its
endpoints, for we can remove any other edge and the graph will
continue to have only convex faces (see Figure \ref{fi:nonessential}).
We denote by $m$ the number of edges of $G$, and by $m_{\text{int}}$ the
number of its interior edges. Recalling properties (i) and (ii) of separable edges, we have $m=m_{\text{int}}+h$ and
\begin{equation} \label{eq:int1}
m_{\text{int}} \le 3v_3 + 2v_{4,2} + v_{4,1} =:a ,
\end{equation}
where $v_3$ is the number of interior vertices of degree $3$ in $G$,
and $v_{4,i}$ is the number of interior vertices
$u$ of degree at least $4$ in $G$ with exactly $i$ edges separable at $u$.
Notice that
\[n=v_3+v_{4,0}+v_{4,1}+v_{4,2}.\]
The estimate in (\ref{eq:int1}) may be pessimistic, because it
doubly counts edges that are separable at both endpoints (such as the one in Figure \ref{fi:nonessential}(c)).
To address this possible over-estimation, denote by $m_{\text{double}}$ the
number of edges that are separable at both endpoints, to which
we refer as {\em doubly separable edges}, and rewrite (\ref{eq:int1}) as
\begin{equation} \label{eq:mInt}
m_{\text{int}} = 3v_3 + 2v_{4,2} +v_{4,1} - m_{\text{double}} = a-m_{\text{double}}.
\end{equation}
Denoting by $f$ the number of bounded faces of
$G$, we have, by Euler's formula, \[n+h+(f+1) = (m_{\text{int}}+h)+2\] (the expression in the parentheses on the left is the number of faces in $G$,
and the expression in the parentheses on the right is the number of edges), or
\begin{equation} \label{eq:mFaces}
f = m_{\text{int}}-n+1.
\end{equation}
Let $f_k$, for $k\ge 3$, denote the number of interior faces of degree $k$ in
$G$. By doubly counting the number of edges in $G$, and then applying (\ref{eq:mFaces}), we get
$$
\sum_{k\ge 3} kf_k = 2m_{\text{int}}+h = 2(f + n -1)+h =
\sum_{k\ge 3} 2f_k + 2n-2+h ,
$$
or
\begin{equation} \label{eq:sumFk}
\sum_{k\ge 3} (k-2)f_k = 2n+h-2 .
\end{equation}
The number of edges that were removed from $T$ is
$\sum_{k\ge 3} (k-3)f_k$, because a face of $G$ of degree $k$
must have had $k-3$ diagonals that were edges of $T$. This number
is therefore

\begin{eqnarray}
\sum_{k\ge 3} (k-3)f_k & = & \sum_{k\ge 3} (k-2)f_k - f = 2n+h-2 - f = \nonumber \\[0.3em]
                       & = & 2n + h - 2 - (m_{\text{int}}-n+1) = 3n+h-3-a+m_{\text{double}} \label{eq:removed}
\end{eqnarray}
(by first applying (\ref{eq:sumFk}), then (\ref{eq:mFaces}), and finally (\ref{eq:mInt})).

\begin{figure}[h]
\centerline{\placefig{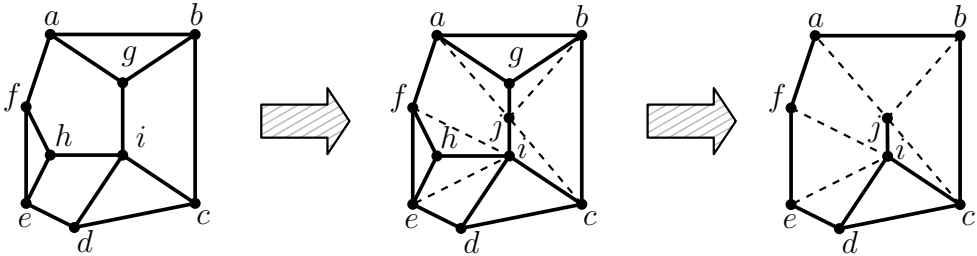}{0.85\textwidth}}
\vspace{-1mm}

\caption{\small \sf A convex decomposition $G$ of $S$, its corresponding graph $G'$, and the reduced form of $G'$ after removing vertices of degree 3.
The edges that have been added are dashed.}
\label{fi:Gprime}
\vspace{-2mm}
\end{figure}

We next derive a lower bound for the right-hand side of (\ref{eq:removed}). For this,
we transform $G$ into another graph $G'$ as follows. We first subdivide each doubly separable edge of $G$ at its midpoint,
say, and add the subdivision point as a new vertex of $G$ (e.g., see the vertex $j$ in Figure \ref{fi:Gprime}). We now
modify $G$ as follows. We take each vertex $u$ of degree $3$
in $G$ and surround it by a triangle, by connecting all pairs of
its three neighbors. Notice that some of these neighbors may be new subdivision vertices, and that
some of the edges of the surrounding triangle may already belong to
$G$. For example, see Figure \ref{fi:Gprime}, where the edges $ei,fi$ are added around the vertex $h$ and the edges $aj,bj$ are added around the vertex $g$.
Next we take each interior vertex $u$ with two separable edges at $u$ and
complete these two edges into a triangle by connecting their other
endpoints, each of which is either an original point or a new subdivision point;
here too the completing edge may already belong to $G$. For example, see the edge $cj$ in Figure \ref{fi:Gprime}, induced by the two separable edges of the vertex $i$.
We then take the resulting graph $G'$ and remove each vertex of degree $3$ and its three incident edges; see the reduced version of $G'$ in Figure \ref{fi:Gprime}.
A crucial and easily verified property of this transformation is
that the newly embedded edges do not cross each other, nor do they
cross old edges of $G$.

The number $f'$ of bounded faces of the new graph $G'$ is at least $v_3+v_{4,2}$,
which is the number of triangles that we have created, and the
number $n'$ of its interior vertices is $n-v_3+m_{\text{double}}$.
Also, $G'$ still has $h$ hull edges. Using Euler's formula, as in (\ref{eq:mFaces}) and (\ref{eq:sumFk}) above, we have
$f' \le 2n'+h-2$. Combining the above, we get
$$
v_3+v_{4,2} \le 2(n-v_3+m_{\text{double}}) +h-2 ,
$$
or
$$
m_{\text{double}} \ge \frac32 v_3 + \frac12 v_{4,2} - n - \frac12 h + 1.
$$
Hence, the right-hand side of (\ref{eq:removed}) is at least \vspace{3mm}

\noindent $\displaystyle 3n+h-3-a+m_{\text{double}}$  \vspace{-2mm}
\begin{eqnarray*}
& \ge & 3n+h-3- \left(3v_3 + 2v_{4,2} + v_{4,1}\right) + \left(\frac32 v_3 + \frac12 v_{4,2} - n - \frac12 h + 1\right) \\[0.4em]
& = & 2n + \frac12 h -2 - \frac32 v_3 - \frac32 v_{4,2} - v_{4,1} \\[0.4em]
& \ge & 2n + \frac12 h - 2 - \frac32 \left( v_3 + v_{4,2} + v_{4,1} + v_{4,0} \right) \\[0.4em]
& = & \frac{n+h}{2} - 2 = \frac{N}{2} - 2 .
\end{eqnarray*}
In other words, the number of edges that we have removed from $T$ is at least
$N/2-2$. On the other hand, we always have $m_{\text{double}} \ge 0$ and $a \le 3n$. Substituting these trivial bounds in (\ref{eq:removed}) we get at least $h-3$ ps-flippable edges.
This completes the proof of the lower bound.

It is easily noticed that only flippable edges of $T$ could have been removed in the initial pruning stage.
Hurtado, Noy, and Urrutia \cite{HNU99} present two distinct triangulations that contain exactly $N/2-2$ flippable edges (one of those is depicted in Figure \ref{fi:lowFlip}).
These triangulations cannot have a set of more than $N/2-2$ ps-flippable edges. Therefore, there are triangulations
for which our bound is tight in the worst case.
Similarly, for point sets in convex position, all $h-3$ interior edges form a set of ps-flippable edges, showing that the other term in the lower bound is also tight in the worst case.
\end{theProof}

\noindent{\bf Remark.}
The proof of Lemma 1.3 actually yields the slightly better bound
$$
\frac12 N + \frac12 v_{4,1} + \frac32 v_{4,0} - 2 .
$$
That is, for the bound to be tight, every interior vertex $u$ of degree $4$ or higher must have two incident
edges separable at $u$ (note that this condition holds vacuously for the triangulation in Figure \ref{fi:lowFlip}).

\paragraph{Convex decompositions.}
The preceding analysis is also related to the notion of {\em convex decompositions}, as defined in the introduction.
Urrutia~\cite{U98} asked what is the minimum number of faces that can always be achieved in a
convex decomposition of any set of $N$ points in the plane. Hosono~\cite{H09}
proved that every planar set of $N$ points admits a convex decomposition with
at most $\lceil\frac{7}{5}(N+2)\rceil$ (bounded) faces. For every $N\geq 4$,
Garc\'{\i}a-Lopez and Nicol\'as~\cite{GL06} constructed $N$-element
point sets that do not admit a convex decomposition with fewer than
$\frac{12}{11}N-2$ faces. By Euler's
formula, if a connected crossing-free straight-edge graph has $N$ vertices and $e$
edges, then it has $e-N+2$ faces (including the exterior face). It
follows that for convex decompositions, minimizing the number of faces
is equivalent to minimizing the number of edges. (For convex decompositions contained in a given triangulation, this is also equivalent to maximizing
the number of removed edges, which form a set of ps-flippable edges.)

Lemma \ref{le:psFlip} directly implies the following corollary. (The bound that it gives is weaker than the bound in \cite{H09},
but it holds for every triangulation.)
\begin{corollary}
Let $S$ be a set of $N$ points in the plane, so that its convex hull has $h$ vertices, and let $T$ be a triangulation of $S$.
Then $T$ contains a convex
decomposition of $S$ with at most $\frac{3}{2}N-h \le \frac{3}{2}N-3$ convex faces and at most $\frac{5}{2}N-h-1 \le \frac{5}{2}N-4$ edges.
Moreover, there exist point sets $S$ of arbitrarily large size, and triangulations $T \in \tr{S}$ for which these bounds are tight.
\end{corollary}

\section{Applications of ps-flippable edges to graph counting} \label{se:apps}
In this section we apply the ps-flippability lemma (Lemma \ref{le:psFlip}) to obtain several improved bounds on the number of crossing-free straight-edge graphs
of various kinds on a fixed set of points in the plane.

\subsection{The ratio between the number of crossing-free straight-edge graphs and the number of triangulations} \label{ssec:pg}

We begin by recalling some observations already made in the introduction.
Let $S$ be a set of $N$ points in the plane.
Every crossing-free straight-edge graph in $\pg{S}$ is contained in at least one triangulation in $\tr{S}$. Additionally, since a triangulation has fewer than $3N$ edges,
every triangulation $T \in \tr{S}$ contains fewer than $2^{3N}=8^N$ crossing-free straight-edge graphs. This immediately implies
\[\pgr{S} < 8^N \cdot \tri{S}.\]
However, this inequality
seems rather weak since it potentially counts some crossing-free straight-edge graphs many times. More formally, given a graph $G \in \pg{S}$ contained in $x$ distinct triangulations of $S$,
we say that $G$ has a \emph{support} of $x$, and write $\supp{G}=x$. Thus, every graph $G \in \pg{S}$ will be counted $\supp{G}$ times in the preceding inequality.

Recently, Razen, Snoeyink, and Welzl \cite{RSW08} managed to break the $8^N$ barrier by overcoming the above inefficiency. However, they obtained only a slight improvement, with the bound
$\pgr{S} = O\left(7.9792^N\right) \cdot \tri{S}$. We now present a more significant improvement, using a much simpler technique that relies on the ps-flippability lemma.

\begin{theorem}
\label{th:triVSpg}
For every set $S$ of $N$ points in the plane, $h$ of which are on the convex hull,
\[ \pgr{S} \le \left\{
\begin{array}{ll}
\displaystyle \frac{(4\sqrt{3})^N}{2^h} \cdot \tri{S} < \frac{6.9283^N}{2^h} \cdot \tri{S}, & \text{\quad for } h \le N/2,\\[3 mm]
8^N \left(3/8\right)^h \cdot \tri{S}, & \text{\quad for } h > N/2.
\end{array} \right. \]
\end{theorem}
\begin{theProof}{\!\!}
The exact value of $\pgr{S}$ is easily seen to be
\begin{equation} \label{eq:pgExact}
\pgr{S} = \sum_{T \in \tr{S}} \sum_{G \in \pg{S} \atop G \subseteq T} \frac{1}{\supp{G}},
\end{equation}
because every graph $G$ appears $\supp{G}$ times in the sum, and thus contributes a total of $\supp{G}\cdot\frac{1}{\supp{G}} = 1$ to the count.
We obtain an upper bound on this sum as follows.
Consider a graph $G \in \pg{S}$ and a triangulation $T \in \tr{S}$, such that $G \subseteq T$. By Lemma \ref{le:psFlip},
there is a set $F$ of $t=\max(N/2-2,h-3)$ ps-flippable edges in $T$.\footnote{Here we implicitly assume that $N$ is even. The case where $N$ is odd is handled in the exact same manner, since a constant change in the size of $F$ does not affect the asymptotic bounds.} Let $F_{\bar{G}}$ denote the set of edges that are in $F$ but
\emph{not} in $G$, and put $j = \left| F_{\bar{G}} \right|$.
Removing the edges of $F_{\bar{G}}$ from $T$ yields a convex decomposition of $S$ which still contains $G$ and whose non-triangular interior faces have a total of $j$ missing diagonals.
Suppose that there are $m$ such faces, with $j_1,j_2,\ldots,j_m$ diagonals respectively, where $\sum_{k=1}^{m}j_k = j$.
Then these faces can be triangulated in $\prod_{k=1}^{m}C_{j_k+1}$ ways, and each of the resulting triangulations contains $G$.
We always have $C_{i+1} \ge 2^i$, for any $i \ge 1$, as is easily verified, and so $\supp{G}\ge 2^j$.
(Equality occurs when all the non-triangular faces of $T\setminus F_{\bar{G}}$ are quadrilaterals.)

Next, we estimate the number of subgraphs $G \subseteq T$ for which the set $F_{\bar{G}}$ is of size $j$. Denote by $E$ the set of edges of $T$ that are not in $F$,
and assume that the convex hull of $S$ has $h$ vertices.
Since there are $3N-3-h$ edges in any triangulation of $S$, $|E| \le 3N-3-h - t$. To obtain a graph $G$
for which $\left| F_{\bar{G}} \right| = j$, we choose any subset of edges from $E$, and any $j$ edges from $F$ (the $j$ edges
of $F$ that will not belong to $G$). Therefore, the number of such subgraphs is at most
$\displaystyle 2^{3N-h-t-3} \cdot \binom{t}{j}$.

We can thus rewrite (\ref{eq:pgExact}) to obtain
\begin{equation*}
\begin{array}{rcl}
\pgr{S} & \le & \displaystyle \sum_{T \in \tr{S}} \sum_{j=0}^{t} 2^{3N-h-t-3} \cdot \binom{t}{j} \cdot \frac{1}{2^j} \\[0.9em]
        & = & \displaystyle \tri{S} \cdot 2^{3N-h-t-3} \sum_{j=0}^{t} \binom{t}{j}\frac{1}{2^j} \\[1.3em]
        & = & \displaystyle \tri{S} \cdot 2^{3N-h-t-3} \cdot (3/2)^{t}.
\end{array}
\end{equation*}
If $t=N/2-2$, we get $\displaystyle \pgr{S} < \tri{S} \cdot \frac{(4\sqrt{3})^N}{2^h} < \frac{6.9283^N}{2^h} \cdot\tri{S}$.
If $t=h-3$, we have $\displaystyle \pgr{S} \le \tri{S} \cdot 2^{3N-2h} \cdot (3/2)^h = \tri{S} \cdot 8^N \cdot (3/8)^h$.
To complete the proof, we note that $N/2-2 > h-3$ when $h \le N/2$.
\end{theProof}

\begin{figure}[h]
\centerline{\placefig{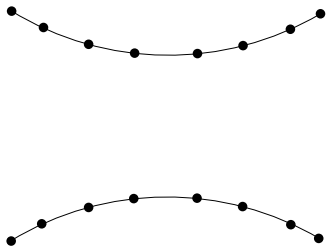}{0.28\textwidth}}
\vspace{-1mm}

\caption{\small \sf A double chain configuration with 16 vertices.}
\label{fi:doubleC}
\vspace{-2mm}
\end{figure}

For a lower bound on $\pgr{S} / \tri{S}$, we consider the \emph{double chain} configurations, presented in \cite{GNT00}
(and depicted in Figure \ref{fi:doubleC}).
It is shown in \cite{GNT00} that, when $S$ is a double chain configuration, $\tri{S}=\Theta^*\left(8^N\right)$ and
$\pgr{S}=\Theta^*\left(39.8^N\right)$ (actually, only the lower bound on $\pgr{N}$ is given in \cite{GNT00}; the upper bound appears in \cite{AHHHKV07}).
Thus, we have $\pgr{S}=\Theta^*\left(4.975^N\right) \cdot\tri{S}$ (for this set $h=4$, so $h$ has no real effect on the asymptotic bound of Theorem \ref{th:triVSpg}).

For another lower bound, consider the case where $S$ is in convex position.
In this case we have $\tri{S} = C_{N-2} = \Theta^*(4^N)$, and $\pgr{S} = \Theta^*(11.65^N)$ (see \cite{FN99}). Hence, $\pgr{S}/\tri{S} = \Theta^*(2.9125^N)$,
whereas the upper bound provided by Theorem \ref{th:triVSpg} is $3^N$ in this case.
Informally, the (rather small) discrepancy between the exact bound in \cite{FN99} and our
bound in the convex case comes from the fact that when $j$ is large, the faces of the resulting convex decomposition are likely to have many edges, which makes
$\supp{G}$ substantially larger than $2^j$.
It is an interesting open problem to exploit this observation to improve our upper bound when $h$ is large.

Finally, recall the notations $\tri{N}=\max_{|S|=N}\tri{S}$ and $\pgr{N}=\max_{|S|=N}\pgr{S}$. Combining the bound $\tri{N} < 30^N$, obtained in \cite{SS10}, with
the first bound of Theorem \ref{th:triVSpg} implies $\pgr{N} < 207.85^N$; see Table \ref{ta:UpperBounds} for comparison with earlier bounds. The bound improves significantly as $h$ gets larger.

\subsection{The number of spanning trees and forests} \label{ssec:st}
\paragraph{Spanning Trees.} For a set $S$ of $N$ points in the plane, we denote by $\st{S}$ the set of all crossing-free straight-edge spanning trees of $S$,
and put $\str{S}:= \left|\st{S}\right|$.
Moreover, we let $\str{N}=\max_{|S|=N}\str{S}$.

Buchin and Schulz \cite{BS10} have recently shown that every crossing-free straight-edge graph contains $O\left(5.2852^N\right)$ spanning trees,
improving upon the earlier bound of $5.\bar{3}^N$ due to Rib\'o Mor and Rote \cite{Rib05,Rot05}.
We thus get $\str{S} = O\left(5.2852^N\right) \cdot \tri{S}$
for every set $S$ of $N$ points in the plane.
The bound from \cite{BS10} cannot be improved much further, since there are triangulations with at least $5.0295^N$ spanning trees \cite{Rib05,Rot05}.
However, the ratio between $\str{S}$ and $\tri{S}$ can be improved beyond that bound, by exploiting and overcoming the same inefficiency as in the case of all crossing-free straight-edge graphs;
that is, the fact that some spanning trees may get multiply counted in many triangulations.

We now derive such an improved ratio by using ps-flippable edges. The proof goes along the same lines of the proof of Theorem \ref{th:triVSpg}.
\begin{theorem}
\label{th:strVtr}
For every set $S$ of $N$ points in the plane,
\[ \str{S} = O\left( 4.7022^N \right) \cdot \tri{S}. \]
\end{theorem}
\begin{theProof}{\!\!}
The exact value of $\str{S}$ is
\[ \str{S} = \sum_{T \in \tr{S}} \sum_{\tau \in \st{S} \atop \tau \subset T} \frac{1}{\supp{\tau}}. \]
Consider a spanning tree $\tau \in \st{S}$ and a triangulation $T \in \tr{S}$, such that $\tau \subset T$. As in Theorem \ref{th:triVSpg},
let $F$ be a set of $N/2-2$ ps-flippable edges in $T$. (Here we do not exploit the alternative bound of $h-3$ on the size of $F$.) Also, let $F_{\bar{\tau}}$ denote the set of edges that are in $F$ but
\emph{not} in $\tau$, and put $j = \left| F_{\bar{\tau}} \right|$. Thus, as argued earlier, $\supp{\tau} \ge 2^j$.

Next, we estimate the number of spanning trees $\tau \subset T$ for which the set $F_{\bar{\tau}}$ is of size $j$.
First, there are $\binom{|F|}{j} < \binom{N/2}{j}$ ways to choose the $j$ edges of $F$ that $\tau$ does not use.
We next contract the $N/2-2-j$ edges of $F$ that were chosen to be in $\tau$ (which will result in having some parallel edges, and possibly also loops) and then remove the remaining edges of $F$.
This produces a non-simple graph $G$ with $N/2+2+j$ vertices and fewer than $5N/2$ edges (recall that by Euler's formula, $G$ contains at most $3N-6$ edges).
Let $S'$ denote the set of vertices of $G$, and let $d_v$ denote the degree in $G$ of a point $v \in S$.
As shown in \cite{BS10,Rib05}, the number of spanning trees in a graph $G$ (not necessarily planar or simple) is at most the product of the vertex degrees in $G$.
Thus, the number of ways to complete the tree is at most
\[ \prod_{v\in S'} d_v \le \left(\frac{\sum_{v \in S'}d_v}{|S'|}\right)^{|S'|} < \left(\frac{5N}{N/2+2+j}\right)^{N/2+2+j}\]
(where we have used the inequality of means for the first inequality).
Hence, there are fewer than $\binom{N/2}{j} \cdot \left(\frac{5N}{N/2+2+j}\right)^{N/2+2+j}$ spanning trees $\tau \subset T$
with $|F_{\bar{\tau}}|=j$. However, when $j$ is large, it is better to use the bound $O\left(5.2852^N\right)$ from \cite{BS10} instead.\footnote{This is not quite correct: When $j$ is close to $N/2$ the former bound is
smaller (e.g., it is $O^*(5^N)$ for $j=N/2$), but we do not know how to exploit this observation to improve the bound.}

We thus get, for a threshold parameter $a<0.25$ that we will set in a moment,
\[ \str{S} < \sum_{T \in \tr{S}} \left( \sum_{j=0}^{aN} \binom{N/2}{j} \cdot \left(\frac{5N}{N/2+2+j}\right)^{N/2+2+j} \cdot \frac{1}{2^j}
+ \sum_{j=aN+1}^{N/2}O\left(5.2852^N\right) \cdot \frac{1}{2^j} \right). \]
The terms in the first sum over $j$ increase when $a \le 0.25$, so the sum is at most $N/2$ times its last term.
Using Stirling's formula, we get that for $a \approx 0.1687$, the last term in the first sum is
$\Theta^*\left(5.2852^N/2^{aN}\right) = O\left(4.7022^N\right)$. Since this also bounds the second sum, we get
\[ \str{S} < \sum_{T \in \tr{S}} O\left(4.7022^N\right) = O\left(4.7022^N\right) \cdot \tri{S}, \]
as asserted. (The optimal parameter $a$ was computed numerically.)
\end{theProof}

Combining the bound just obtained with $\tri{N} < 30^N$ \cite{SS10} implies
\begin{corollary}
$\str{N} = O\left(141.07^N \right)$.
\end{corollary}
This improves all previous upper bounds, the smallest of which is $O(158.6^N)$  \cite{BS10,SS10}. \vspace{2mm}

\noindent {\bf Remark.} It would be interesting to refine the bound in Theorem \ref{th:strVtr} so that it also depends on $h$, as in Theorem \ref{th:triVSpg}.
An extreme situation is when $S$ is in convex position (in which case $|F|=N-3$). In this case it is known that
$\tri{S}=\Theta^*(4^N)$ and $\str{S}=\Theta^*(6.75^N)$ (see \cite{FN99}), so the exact ratio is only $\str{S} / \tri{S} = \Theta^*(1.6875^N)$.
This might suggest that when $h$ is large the ratio should be considerably smaller, but we have not pursued this in this paper.

\paragraph{Forests.} For a set $S$ of $N$ points in the plane, we denote by $\f{S}$ the set of all crossing-free straight-edge forests
(i.e., cycle-free graphs) of $S$, and put $\fr{S}:= \left|\f{S}\right|$.
Moreover, we let $\fr{N}=\max_{|S|=N}\fr{S}$. Buchin and Schulz \cite{BS10} have recently shown that every crossing-free straight-edge graph contains $O\left(6.4884^N\right)$ forests
(improving a simple upper bound of $O^*(6.75^N)$ observed in \cite{AHHHKV07}).
Following the approach of~\cite{BS10}, we combine the bounds for spanning trees (just established) and for plane graphs with a bounded number of edges (established in Section \ref{ssec:restrict} below), to obtain the following result.
\begin{theorem}
\label{th:frVStr}
For every set $S$ of $N$ points in the plane,
\[ \fr{S} = O\left(5.3514^N\right) \cdot \tri{S}. \]
\end{theorem}
\begin{theProof}{\!\!}
We define a $k$-\emph{forest} to be a forest that has $k$ connected components. The number of $k$-forests of a set $S$  is denoted by $\kfor{S}{k}$.
Since any spanning tree has $N-1$ edges, every $k$-forest has $N-k$ edges.
One way to bound $\kfor{S}{k}$ is by counting the number of plane graphs with $N-k$ edges. This number is bounded in Theorem~\ref{th:triVSpgc} (from the following subsection), where the parameter $c$ in that theorem is equal to $1-k/N$; let us denote this bound as $g_1(N,k)$.
On the other hand, every $k$-forest is obtained by deleting $k-1$ edges from a spanning tree. This allows us to bound the number of $k$-forests in terms of $\str{S}$.
Using Theorem~\ref{th:strVtr} we get the bound $\kfor{S}{k}\le\binom{N-1}{k-1} \cdot O\left( 4.7022^N \right) \cdot \tri{S}$; denote this bound as $g_2(N,k)$. To bound $\fr{S}$, we evaluate $\max_k \min\{g_1(N,k),g_2(N,k)\}$.
A numerical calculation shows that  the maximum value is obtained for $k'\approx 0.0285N$, and the theorem follows since   $\min\{g_1(N,k'),g_2(N,k')\}= O\left (5{.}3514^N \right) \cdot \tri{S}$.
\end{theProof}
As in the previous cases, we can combine this with the bound $\tri{N} < 30^N$ \cite{SS10} to obtain
\begin{corollary}
$\fr{N} = O\left(160.55^N \right)$.
\end{corollary}
Again, this should be compared with the best previous upper bound $O(194.7^N$) \cite{BS10,SS10}.

Consider once again the case where $S$ consists of $N$ points in convex position. In this case we have $\tri{S}=\Theta^*(4^N)$ and $\fr{S}=\Theta^*(8.22^N)$ (see \cite{FN99}),
so the exact ratio is $\fr{S} / \tri{S} = \Theta^*(2.055^N)$, again suggesting that the ratio should be smaller when $h$ is large.

\subsection{The number of crossing-free straight-edge graphs with a bounded number of edges} \label{ssec:restrict}
In this subsection we derive  upper bounds for the number of crossing-free straight-edge graphs on a set $S$ of $N$ points in the plane, with some constraints on the number of edges.
Specifically, we bound the number of crossing-free straight-edge graphs with exactly $cN$ edges, with at most $cN$ edges, and with at least $cN$ edges.
The first variant has already been used in the preceding subsection for bounding the number of forests.

\paragraph{Crossing-free straight-edge graphs with exactly  \mbox{\boldmath $cN$} edges.}



We denote by $\pgc{S}{c}$ the set of all crossing-free straight-edge graphs of $S$ with exactly $cN$ edges, and put $\pgrc{S}{c}:= \left|\pgc{S}{c}\right|$.
The following theorem, whose proof goes along the same lines of the proof of Theorem \ref{th:triVSpg}, gives a bound for $\pgrc{S}{c}$
 \begin{theorem}
\label{th:triVSpgc}
For every set $S$ of $N$ points in the plane and $0 \leq c < 3$,
\[\pgrc{S}{c} =O^*\left(B(c)^N\right) \cdot \tri{S}, \]
where
\begin{align*}
B(c)&:=  \frac{5^{5/2}}{8(c+t-1/2)^{c+t-1/2} (3-c-t)^{3-c-t}(2t)^t(1/2-t)^{1/2-t}},
\end{align*}
and
\begin{equation} \label{eq:t}
t = \frac{1}{2} \left(\sqrt{(7/2)^2 + 3c + c^2 } - 5/2 - c \right).
\end{equation}
\end{theorem}
See Figure \ref{fi:plot1} for a plot of the base $B(c)$ as a function of $c$.\vspace{2mm}

\begin{theProof}{\!\!}
The exact value of $\pgrc{S}{c}$ is
\[ \pgrc{S}{c} = \sum_{T \in \tr{S}} \sum_{G \in \pgc{S}{c} \atop G \subseteq T} \frac{1}{\supp{G}}, \]
where $\supp{G}$, the support of $G$, is defined as in the case of general crossing-free straight-edge graphs treated in Section \ref{ssec:pg}.
We obtain an upper bound on this sum as follows.
Consider a graph $G \in \pgc{S}{c}$ and a triangulation $T \in \tr{S}$, such that $G \subseteq T$. By Lemma \ref{le:psFlip},
there is a set $F$ of $N/2-2$ ps-flippable edges in $T$. Let $F_{\bar{G}}$ denote the set of edges that are in $F$ but
\emph{not} in $G$, and put $j = \left| F_{\bar{G}} \right|$. As in the preceding proofs, we have $\supp{G} \ge 2^j$.

Next, we estimate the number of subgraphs $G \subseteq T$ for which the set $F_{\bar{G}}$ is of size $j$.
Denote by $E$ the set of edges of $T$ that are not in $F$.
As argued above, $|E| < 5N/2$. To obtain a graph
for which $\left| F_{\bar{G}} \right| = j$, we choose any $j$ edges from $F$ (the $j$ edges
of $F$ that will not belong to $G$), and any subset of  $cN - \left(N/2-2 - j \right) = (c-1/2)N + j + 2$ edges from $E$.
If $(c-1/2)N + j + 2 < 0$, there are no such graphs and we ignore these values of $j$. The number of ways to pick the edges from $E$ is at most
$
 O^*\left( \binom{5N/2}{(c-1/2)N + j }\right).
$
This implies that
\begin{eqnarray}
\pgrc{S}{c} & < & \displaystyle \sum_{T \in \tr{S}} \sum_{j=0}^{N/2} O^*\left(\binom{5N/2}{(c-1/2)N + j} \cdot \binom{N/2}{j}\right) \cdot \frac{1}{2^j} \nonumber \\[0.3em]
        & = & \displaystyle \tri{S} \cdot \sum_{j=0}^{N/2} O^*\left(\binom{5N/2}{(c-1/2)N + j} \cdot \binom{N/2}{j}\right) \cdot \frac{1}{2^j}. \label{eq:atMost}
\end{eqnarray}
(As already noted, when $c<1/2$, only the terms for which $(c-1/2)N + j \ge 0$ are taken into account.)

As in the preceding subsection, it suffices to consider only the largest term of the sum. For this, we consider the quotient of the $j$-th and $(j-1)$-st terms (ignoring the $O^*(\cdot)$ notation, which
will not affect the exponential order of growth of the terms), which is
\[ \displaystyle \frac{ \binom{N/2}{j} \binom{5N/2}{(c-1/2)N+j} }{ 2\binom{N/2}{j-1} \binom{5N/2}{(c-1/2)N+j-1} } =
\frac{\Big(N/2 - j +1\Big)\Big(5N/2 - (c - 1/2)N - j + 1\Big) }{2j\Big((c-1/2)N+j\Big)  }. \]
To simplify matters, we put $a = N/2$ and $b = (c-1/2)N$. Moreover, since we are only looking for an asymptotic bound,
and are willing to incur small multiplicative errors within the $O^*(\cdot)$ notation, we may ignore the two $+1$ terms in the numerator
when $N$ is sufficiently large; we omit the routine algebraic justification of this statement.
The above quotient then becomes
(approximately) $ \displaystyle \frac{(a - j)(5a - b - j) }{2j(b+j) }$, which is larger than 1 whenever
\begin{equation*}
\begin{array}{rcl}
j & < & \frac{1}{2} ( \sqrt{ 56a^2 + 8ab + b^2 } - 6a - b ) = \\[1.2em]
  & = & \frac{N}{2} \left(\sqrt{(7/2)^2 + 3c + c^2} - 5/2 - c \right) = tN,
\end{array}
\end{equation*}
with $t$ given in (\ref{eq:t}). A simple calculation shows that $0 \le t < 1/2$ and $0 \le c-1/2+t \le 5/2$ for $0 \le c \le 3$.
In other words (and rather unsurprisingly), the index $j=tN$ attaining the maximum does indeed lie in the range where the two binomial coefficients in the corresponding terms in (\ref{eq:atMost})
are both well defined (non-zero).

Now that we have the largest term of the sum in (\ref{eq:atMost}), we obtain
\[ \pgrc{S}{c} = \tri{S} \cdot O^*\left( \binom{5N/2}{(c-1/2)N + tN} \cdot \binom{N/2}{tN} \cdot \frac{1}{2^{tN}} \right). \]
Using Stirling's approximation, we have
\begin{equation*}
\begin{array}{rcl}
\pgrc{S}{c} & = & \displaystyle \tri{S} \cdot O^*\left( \left( \frac{(5/2)^{5/2}}{(c+t-1/2)^{c+t-1/2} (3-c-t)^{3-c-t}} \cdot \frac{(1/2)^{1/2}}{t^t (1/2-t)^{1/2-t}} \cdot \frac{1}{2^t} \right)^N \right) \\[1.2em]
            & = & \displaystyle \tri{S} \cdot O^*\left( \left( \frac{5^{5/2}}{8(c+t-1/2)^{c+t-1/2} (3-c-t)^{3-c-t}(2t)^t(1/2-t)^{1/2-t}} \right)^N \right),
\end{array}
\end{equation*}
as asserted.
\end{theProof}

%
%
%
%
%
\paragraph{Crossing-free straight-edge graphs with at most \mbox{\boldmath $cN$} edges.}
For a set $S$ of $N$ points in the plane and a constant $0< c < 3$, we denote by $\pgcc{S}{c}$ the set of all crossing-free straight-edge graphs of $S$ with at most $cN$ edges, and put $\pgrcc{S}{c}:= \left|\pgcc{S}{c}\right|$. The bound for  $\pgrc{S}{c}$ in Theorem~\ref{th:triVSpgc} helps us to determine the bound for $\pgrcc{S}{c}$.
 \begin{theorem}
\label{th:triVSpgcc}
For every set $S$ of $N$ points in the plane and $0< c < 3 $,
\[ \pgrcc{S}{c}  = \begin{cases} O^*\left(B(c)^N\right) \cdot \tri{S} & \text{if $c\le 19/12$},
\\ O^*\left((4\sqrt{3})^{N}\right) \cdot \tri{S} & \text{otherwise},\end{cases}
\]
where $B(c)$ is defined as in Theorem \ref{th:triVSpgc}.
\end{theorem}
\begin{theProof}{\!\!}
We begin by noticing that
\begin{align}\label{eq:simplesum}
\pgrcc{S}{c}=\sum_{0< j\le cN} \pgrc{S}{j/N}= O^*\left(\max_{c'\le c} \pgrc{S}{c'} \right)= O^*\left(\max_{c'\le c} B(c')^N \right) \cdot \tri{S}.
\end{align}
Let $F(c)$ be the natural logarithm of the non-constant part of the denominator of $B(c)$ (the numerator is a constant). That is,
\[ F(c) = (c+t-1/2)\ln(c+t-1/2) + (3-c-t)\ln(3-c-t) +t\ln(2t) +(1/2-t)\ln(1/2-t). \]
Since each of the terms in the logarithms is positive when $0<c<3$, finding a maximum for $B(c)$ in this range is equivalent to finding a minimum for $F(c)$.
Put $t'=t'(c)$ (the derivative of $t$ as a function of $c$). Then
\begin{align*}
F'(c) &= (1+t')\ln(c+t-1/2) +(1+t') -(1+t')\ln(3-c-t)-(1+t') \\[0.5em]
&+ t'\ln(2t) + t' -t'\ln(1/2-t) -t' \\[0.5em]
&= (1+t')\ln\left(\frac{c+t-1/2}{3-c-t}\right) +t'\ln\left(\frac{2t}{1/2-t} \right)
\end{align*}
For $c=19/12$ we have $t=1/6$ and the arguments of both logarithms are 1, so $F'(c)=0$.
Easy calculations show that
\[t' = \frac{1}{2} \left(\frac{3/2+c}{\sqrt{(7/2)^2 + 3c + c^2 }} - 1 \right)= -\frac{0.5+t}{\sqrt{(7/2)^2 + 3c + c^2} } .\]
This is easily seen to imply that $t'\le 0$ and $1+t'\ge 0$ for $0<c<3$.
This implies that $c+t$ is monotone increasing with $c$, and that $t$ is decreasing (because $1+t'\ge 0$ and $t'\le 0$). It follows that $F'(c)$ is increasing with $c$, implying that $F(c)$ attains its minimum at $c=19/12$ (since $F'(19/12)=0$).
Another easy calculation shows that $B(19/12)=4\sqrt{3}$.
\end{theProof}

For example, Theorem \ref{th:triVSpgcc} implies that there are at most $O^* \left(5.4830^N\right) \cdot \tri{S}$ crossing-free straight-edge graphs with at most $N$ edges, over any set $S$ of $N$ points in the plane.
In particular, this is also an upper bound on the number of crossing-free straight-edge forests on $S$, or of spanning trees,
or of spanning cycles. Of course, better bounds exist for these three special cases, as demonstrated earlier in this paper for the first two bounds.

\paragraph{Crossing-free straight-edge graphs with at least \mbox{\boldmath $cN$} edges.}
We next bound the number of plane graphs with at least $cN$ edges. Following the notations used above, we denote by $\pgccc{S}{c}$ the set of all crossing-free straight-edge graphs of $S$ with at least $cN$ edges, and put $\pgrccc{S}{c}:= \left|\pgccc{S}{c}\right|$.

 \begin{theorem}
\label{th:triVSpgccc}
For every set $S$ of $N$ points in the plane and $0< c < 3 $,
\[ \pgrccc{S}{c}  = \begin{cases} O^*\left(B(c)^N\right)  \cdot \tri{S}  & \text{if $c\ge 19/12$},
\\ O^*\left((4\sqrt{3})^N\right) \cdot \tri{S} & \text{otherwise}.\end{cases}
\]
\end{theorem}
\begin{theProof}{\!\!}
Similar to Equation~\eqref{eq:simplesum} we can bound $\pgrccc{S}{c}$ by
\begin{align}\label{eq:simplesum2}
\pgrccc{S}{c}=\sum_{cN \le j < 3N} \pgrc{S}{j/N}= O^*\left(\max_{c \le c' < 3} \pgrc{S}{c'} \right)= O^*\left(\max_{c \le c' < 3} B(c)^N \right)  \cdot \tri{S} .
\end{align}
The analysis of \eqref{eq:simplesum2} is symmetric to the one presented in the proof of Theorem \ref{th:triVSpgcc},
and the theorem follows.
\end{theProof}

\begin{figure}[h]
\centerline{\placefig{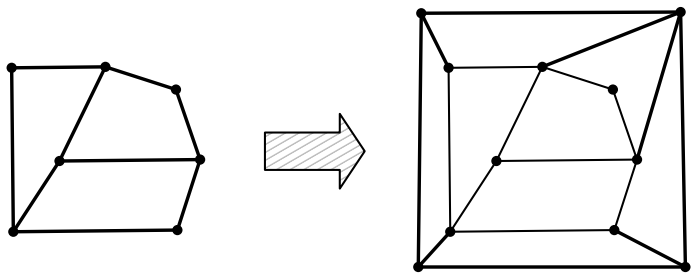}{0.6\textwidth}}
\vspace{-1mm}

\caption{\small \sf A quadrangulation of $S$ and a quadrangulation of $S'$ that contains it.}
\label{fi:quadrangulations}
\vspace{-2mm}
\end{figure}

As an application, consider the problem of bounding the number of quadrangulations of $S$, namely crossing-free straight-edge connected graphs
on the vertex set $S$ with no isolated vertices,
that include all the hull edges of $conv(S)$, and where every bounded face is a quadrilateral.
When $h$ is odd, no quadrangulations can be embedded over $S$ (e.g., see \cite{BT95}).
We may thus assume that $h$ is even, and create a superset $S' \supset S$ as follows.
We take a quadrilateral $Q$ that contains $S$ in its interior, and add the vertices of $Q$ to $S$.
It is easy to see that every quadrangulation of $S$ is contained in at least one quadrangulation of $S'$; see Figure \ref{fi:quadrangulations} for an illustration\footnote{
We need to construct a quadrangulation of the annulus-like region between $Q$ and the convex hull of $S$. We start by connecting a vertex of $Q$ to a
vertex of the convex hull, and in each step we add a quadrangle by either marching along two edges of the hull or along one edge of the hull and one  edge of $Q$.
This produces the desired quadrangulation.}.
Therefore, it suffices to bound the number of quadrangulations of $S'$.

Using Euler's formula, we notice that a quadrangulation of $S'$ has $N+1$ quadrilaterals, and $2N+4$ edges (since $|S'|=N+4$).
Therefore, we can use Theorem \ref{th:triVSpgcc} with $c=2$, which implies a bound of $O^*(6.1406^N)\cdot \tri{N} = O(184.22^N)$. (There are actually $N+4$ points
and $c=(2N+4)/(N+4)$. However, since we are only
interested in the exponential part of the bound, the above bound, with the $O^*(\cdot)$ notation, does hold.)
We are not aware of any previous explicit treatment of this problem.

\section{Conclusion}
In this paper we have introduced the notion of pseudo-simultaneously flippable edges in triangulations, have shown that many such edges always exist,
and have used them to obtain several refined bounds
on the number of crossing-free straight-edge graphs on a fixed (labeled) set of $N$ points in the plane.
The paper raises several open problems and directions for future research.

One such question is whether it is possible to further extend the notion of ps-flippability. For example, one could consider, within a fixed triangulation $T$,
 the set of diagonals of a collection of pairwise interior-disjoint simple, but not necessarily convex, polygons. The number of such
 diagonals is likely to be larger than the size of the maximal set of ps-flippable edges, but it not clear how large is the number of triangulations
 that can be obtained by redrawing diagonals.

We are currently working on two extensions to this work. The first extends our techniques to the cases of crossing-free straight-edge perfect matchings and spanning (Hamiltonian) cycles.
This is done within the linear-algebra framework introduced by Kasteleyn (see \cite{BKKSS07,LP86}), and can be found in \cite{SSW11}. The second work studies charging schemes in which the charge is moved
across certain objects belonging to different crossing-free straight-edge graphs over the same point set. This cross-graph charging scheme allows us to obtain bounds that do not
depend on the current upper bound for $\tri{N}$ (and bounds that depend on $\tri{N}$ in a non-linear fashion).


\end{document}